\documentclass[12pt,single,a4,epsfig]{article}
\usepackage{graphicx}
\usepackage{epsfig}
\usepackage{amssymb}
\usepackage{latexsym}
\usepackage{revsymb}
\begin{document}
\baselineskip=6mm
\date{}
\title{Anisotropy in spatial order-disorder transformations and the vortex lattice symmetry transition in $YNi_2B_2C$ and $LuNi_2B_2C$}
\maketitle
\vspace{-2cm}
\author{D. Jaiswal-Nagar$^{1,2,3*}$, T. Isshiki$^1$, N. Kimura$^1$, H. Aoki$^1$, H. Takeya$^4$, S. Ramakrishnan$^2$\\ and A.K. Grover$^2$}\\
$^1$Centre for Low Temperature Science, Tohoku University, Sendai, Miyagi, 980-8578, Japan\\ $^2$DCMP$\&$MS, Tata Institute of Fundamental Research, Mumbai-400005, India\\ $^3$Physikalishes Institut, Goethe-Universit$\ddot{a}$t, 60438 Frankfurt am Main, Germany\\ $^4$National Institute for Materials Science, Sengen 1-2-1, Tsukuba, Ibaraki 305-0047, Japan\\
$^*$e-mail:Jaiswal-Nagar@physik.uni-frankfurt.de
%\date{\today}
\maketitle

\noindent

\abstract
{Explorations of the order-disorder transformation in vortex matter in single crystals of tetragonal structured (c/a~ $\sim$~3) borocarbide superconductors, $YNi_2B_2C$ and $LuNi_2B_2C$, reveal that vortex arrays experience different effective pinning in different crystallographic directions. We surmise that correlation exists between the large anisotropy in effective pinning/disorder and the differences in the (local) symmetry transition from rhombohedral to (quasi) square vortex lattice(VL). For field along high symmetry directions, like, c-axis and ab-basal plane, the VL symmetry is close to square and the ordered state spans a large field interval. When the field is turned away from the c-axis towards ab-plane, at intermediate angles, the region of ordered state shrinks, in response to enhancement in effective pinning. At such intermediate angles the symmetry of the VL would be far from ideal triangular or square.}

\section*{Introduction}

 Disorder can play an instructive role in unraveling important aspects of phase transitions in a condensed matter system. In the context of vortex arrays, the role of two kinds of disorder, namely, thermal fluctuations and quenched random disorder (QRD), in enunciating different phases of vortex matter is well documented \cite{blatter,g-l,shobho,ling,khaykovich,nishizaki}. Quaternary borocarbide superconductors have emerged as convenient test beds to explore a variety of newer issues in vortex physics, e.g., field induced changes in the symmetry of the vortex lattice (VL)\cite{canfield,eskildsen1,mckpaul,dewhurst,vinnikov1,vinnikov2,sakata,eskildsen2}, symmetry changes due to interplay between non-local effects in electrodynamics and symmetry of the superconducting order parameter \cite{park,kogan,gammel,nakai}, etc. From angle dependent dc magnetisation hysteresis and ac shielding response measurements in crystals of two borocarbides, viz., $Y(Lu)Ni_{2}B_{2}C$ (Y(L)NBC), we show that the effective disorder experienced by the VL is different in different directions, which reflects in the details of their vortex phase diagrams. In particular, we conjecture that the closer the VL conforms to the perfect square/triangle symmetry, larger is the correlation volume for the ordered VL and it spans a larger portion of vortex phase space.\\
Our study is based on a premise that a transformation/transition amongst vortex phases can, {\it prima facie}, be captured via an anomalous change in the pinning attribute of the vortices, as evidenced in the field-temperature (H,T) dependence of the critical current density $J_c$ of a given superconductor. We obtain information on $J_c$(H,T) via an exploration of the peak effect (PE) phenomenon (typically occurring near/prior to the normal-superconductor phase boundary \cite{shobho,ling,nishizaki}) and a second magnetization peak (SMP) anomaly \cite{khaykovich,nishizaki} lying deeper in the mixed state of weakly pinned single crystals of YNBC and LNBC. Both these compounds are prime examples of field induced transitions in the local symmetry of the VL. For H $||$ c, VL in them \cite{eskildsen1,mckpaul,dewhurst,vinnikov1,vinnikov2,sakata,eskildsen2}, initially undergoes a first order transition from a rhombus symmetry of one kind ($\beta < 60^{\circ}$) to another ($\beta > 60^{\circ}$) via a $45^{\circ}$ reorientation, which eventually becomes a square symmetry via a second order transition with progressive increase in field \cite{eskildsen1,mckpaul,dewhurst,vinnikov1,vinnikov2}. For H $\perp$ c, however, the first order transition is via a $90^{\circ}$ reorientation and the rhombus ($\beta > 60^{\circ}$) never reaches a square symmetry (saturates with $\beta \approx 82^{\circ}$) \cite{sakata,eskildsen2}. From angle dependent PE measurements, we find that the details of the symmetry transitions in the VL in different orientations result in deciding the quality of spatial order in VL. The orientation of the magnetic field in directions along which the VL probably does not assume a symmetric shape (triangle/square with $\beta \rightarrow 60^{\circ}/90^{\circ}$, respectively), the vortex state is only partially ordered, and it eventually amorphizes in multiple steps via the occurrence of SMP and/or multiple magnetization peaks (MMP) apart from the PE. To fortify our assertion that the anomalies prior to the PE are (purely) disorder induced, we enhance the driving effect of the ac field on the partially ordered VL to improve its quality, which in turn suppresses the occurrence of anomalies precursor to PE.

\section*{Experimental details}
The YNBC crystal ($T_c \approx$ 15~K), grown by using an infrared image furnace \cite{takeya}, was cut as parallelopiped (see Fig.~1(b)) with dimension l = 3~mm, b = 0.7~mm, t = 0.67~mm. Its largest dimension is parallel to the a-axis of YNBC. The LNBC crystal on the other hand, was grown by the flux method \cite{canfield}; it is in the form of a thin platelet, with the c-axis perpendicular to its plane. The ac susceptibility measurements were performed on a top loading dilution refrigerator at the Centre for Low Temperature Science, Tohoku University, with amplitudes of the ac field ranging from 0.5 mT to 10 mT (peak to peak). Before measuring the ac susceptibility response, the sample was aligned along a particular crystallographic direction with the help of Laue diffraction method, within an accuracy of $\pm$ $2^{\circ}$. For angular dependent studies, the crystal was rotated (i) from [100] to [110] in the (001) plane, (ii) from [110] to [001] and vice-versa in the (1$\bar{1}$0) plane and (iii) from [001] to [100] and vice-versa in the (010) plane. It was noted that observed response at a given angle, was independent of the phase of the rotation. DC magnetization measurements were made using a commercial 12 Tesla Vibrating Sample Magnetometer (Oxford Instruments, U.K.) at the Tata Institute of Fundamental Research. The upper critical fields, $H_{c2}$, for LNBC crystal at T $\sim$~0.1 K, for H $||$ [001], [110] and [100] are estimated to be 7.4~T, 7.9~T and 8.7~T, respectively. Each of these values corresponds to an ac susceptibility response merging into background.

\section*{Results and discussion}
The inset panel of Fig.~1 (a) shows a M-H hysteresis loop in YNBC at 2.06~K with H $||$ c. The main panel of Fig.~1 (a) depicts the data on an expanded scale (blue curve for H $||$ c and black curve for H $\perp$ c ) to focus on the PE anomaly seen as a hysteresis ``bubble''. It can be noted that while the magnetisation hysteresis widths ($\propto \Delta$M (H)) are notionally comparable in the disordered PE region for both the orientations, they are very different in the field range prior to the onset of the PE (H $<$ $H_{on}$); $\Delta$M is negligibly small for H $||$ c as compared to that for H $\perp$ c. Recall that $\Delta$M (H) $\propto$ $J_c$ (H) \cite{bean,fietz}, which in turn $\propto$ $\frac{1}{\sqrt{V_c}}$ \cite{lo}, where $V_c$ denotes the correlation volume over which the VL is well correlated. The above observation then, implies that a better ordered VL (with a large $V_c$) is formed when H $||$ c, as compared to that for H $\perp$ c. This finding is significant as the superconducting parameters of tetragonal structured (c/a~$\sim$~3) Y(L)NBC show little anisotropy (1~$\sim$~1.3) \cite{hc1,me,metlushko}.
\begin{figure}[!htb]
\begin{center}
\includegraphics[scale=0.3,angle=0] {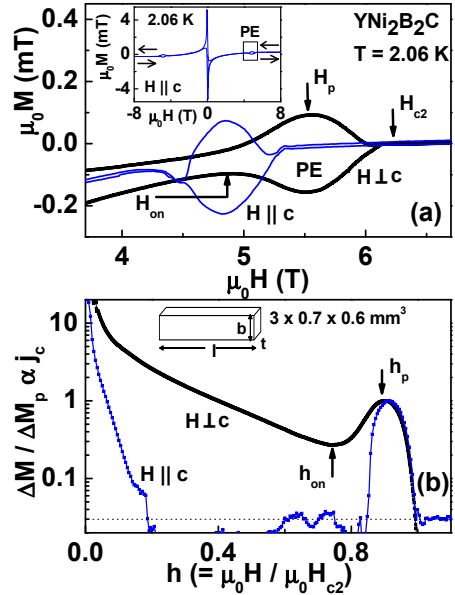}
\caption{ (Color online) (a) Main panel depicts a portion of the M-H hysteresis loop in a parallelopiped crystal of $YNi_2B_2C$ for H $\perp$ c (black) and for H $||$ c (blue), while the inset shows the full M-H loop for H $||$ c. The onset field  $H_{on}$, the peak field $H_p$ of the peak effect, as well as the upper critical field (at which the superconductor becomes normal) $H_{c2}$, have been marked for H $\perp$ c. (b) Normalized hysteresis width ($\propto J_c$) vs. reduced field h = $\mu_0H/\mu_0H_{c2}$, corresponding to the M-H plots in panel (a). The dotted line near zero $j_c$ is a guide to the eye, it represents the experimental accuracy in the measured hysteresis width.}
\label{fig.1}
\end{center}
\end{figure}

During exploration of the PE phenomenon in spherical as well as an irregular shaped single crystals of a cubic superconductor $V_3$Si, K$\ddot{u}$pfer {\it et al.} \cite{kupfer} found that, {\it prima facie}, the disordered state of the VL had no angular dependence for H $> H_p$, while the state of order before the onset of PE crucially depends on the symmetry of the field direction with respect to the underlying CL. Taking a cue from this observation, we show in Fig.~1 (b), the plots of $\Delta$M normalized to its value at peak field $H_p$, versus reduced field, h (= $\mu_0$H/$\mu_0H_{c2}$) for H $||$ c and H $\perp$ c. It is satisfying to note that both the curves merge for H $>$ $H_p$ (similar to those in the $V_3$Si case \cite{kupfer}), thereby elucidating the insensitivity of the disordered vortex matter to the field-direction, for H $>$ $H_p$, in YNBC as well.\\
Considering that the superconducting parameters (coherence length $\xi_c$/$\xi_{ab}$, penetration depth $\lambda_c$/$\lambda_{ab}$, etc.) show little anisotropy ($\sim$ 1-1.2) for rotation from the c-axis to the basal ab-plane directions \cite{hc1,me,metlushko}, the observation that $\frac {({\Delta M_{H_p}/\Delta M_{H_{on}})}_{H||c}}{({\Delta M_{H_p}/\Delta M_{H_{on}})}_{H \perp c}} \approx$~17, suggests that the behaviour of the spatial order of the VL is not governed by the intrinsic superconducting anisotropy of YNBC. It is fruitful at this juncture to recall reports \cite{eskildsen1,sakata,eskildsen2} that a square lattice gets formed for H $||$ c, whereas only a quasi-square lattice (apex angle $\approx 82^{\circ}$) is observed for H $\perp$ c. This observation along with the inference of a larger correlation volume for ordered VL for H $||$ c compared to that for H $\perp$ c, then implies that a better spatially ordered VL is obtained, when its underlying symmetry is a perfect square rather than a quasi-square.\\
In order to obtain information on the spatial ordering in the VL in directions other than H $||$ c and H $\perp$ c, we took a recourse to ac susceptibility measurements in LNBC, on which dHvA measurements were already being performed at 0.1~K \cite{toshi}. In the main panel of Fig.~2 we show angle dependent plots of normalised in-phase susceptibility $\chi^{\prime}_N$ (= $\chi^{\prime}$(H)/$\chi^{\prime}$($H_p$)) at T =~0.1~K with an $h_{ac}$ of 1~mT (peak to peak) at 67~Hz. The normalisation for this as well as subsequent measurements has been chosen in accordance with the discussion above. The PE feature is clearly observable for all the angles in Fig.~2. Note first that prior to the PE region (H $< H_{on}$), $\chi^{\prime}_N$ ($\propto J_c$) \cite{budnick} is the smallest for H $||$ c (i.e., [001]) as compared to the other orientations, suggesting that the VL of LNBC also has better spatial order for H $||$ c as compared to any of the directions within the ab-basal plane. Also, note that within the basal plane, for H $< H_{on}$, $\chi^{\prime}_N$ is least for H $||$ [110] and highest for H $||$ [100], thereby indicating that the VL for H $||$ [110] is better spatially ordered than that for H $||$ [100], reminiscent of the behaviour reported in $V_3$Si \cite{kupfer}.\\
The inset panel in Fig.~2 shows plots of $\chi^{\prime}_N$ vs. reduced field h for rotation of the field from [100] to [110]. In terms of h, the onset field of the PE ($h_{on}$), increases from $\sim$~0.65 for H $||$ [100] to a value of $\sim$~0.72 for H $||$ [110]. A premise, that $h_{on}$ is governed by the residual disorder in the ordered state of the VL, (i.e., higher the residual disorder, lower is the $h_{on}$ and vice-versa) \cite{kupfer2}, then, suggests that the vortex state for H $||$ [110] is not only better ordered as compared to any other orientation in the ab-basal plane, but the ordered state also spans a larger field interval prior to the PE. This finding is significant as the intrinsic superconducting parameters of the tetragonal structured (c/a~$\sim$~3) Y(L)NBC compounds show little anisotropy (1~$\sim$~1.2) \cite{hc1,me,metlushko}.
\begin{figure}[!htb]
\begin{center}
\includegraphics[scale=0.3,angle=270] {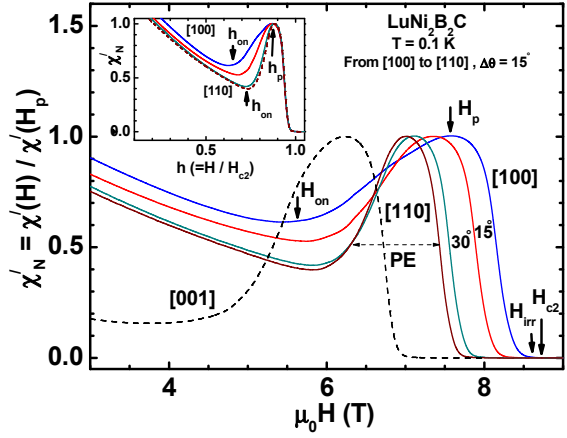}
\caption{ (Color online) The solid curves in the main panel show the angular dependence of normalised in-phase ac susceptibility in $LuNi_2B_2C$ at T =~0.1~K for the in plane rotation of the field from [100] to [110] in increments of $15^{\circ}$, while the black dotted curve represents the $\chi^{\prime}_N$ for H $||$ [001]. The onset field of PE $H_{on}$, the peak field $H_p$, the irreversibility field $H_{irr}$, and the upper critical field $H_{c2}$ have been marked for H $||$ [100] (blue curve). The inset shows $\chi^{\prime}_N$ vs. h (= $\mu_0$H/$\mu_0H_{c2}$) for the rotations in the basal plane from [100] to [110].}
\label{fig. 2}
\end{center}
\end{figure}

We now recall that Park {\it et al.} \cite{park} had surmised that hexagonal to (quasi) square transition for VL in LNBC occurs at a lower threshold field for H $||$ [110] as compared to that for H $||$ [100]. This could imply that well ordered Bragg glass (BG) like phase \cite{g-l} with no dislocations and having (quasi) square symmetry could permeate the entire sample at a lower field for H $||$ [110] as compared to that for H $||$ [100]. The occurrence of symmetry re-orientation transition at a lower field value facilitates the extension of spatial correlations to a larger volume. The symmetry transition in a VL is understood to arise due to a correlation of the VL with the underlying CL \cite{kogan}. A non-local relationship between the super-current and the vector potential results in a distortion of the super-current, which leads to an anisotropic intervortex interaction. The threshold field at which non-local relationship governs the symmetry transitions in VL for H $||$ c has been shown to enhance as the electron mean free path due to impurity effect progressively decreases in Lu$(Ni_{1-x}Co_x)_2B_2C$ series of alloys \cite{gammel}. If such a notion is extended to hold for the state of order of the VL for H $\perp$ c, it would in turn confirm that the effective disorder being experienced for H $||$ [110] is less than that for H $||$ [100] in LNBC.
\begin{figure}[!htb]
\begin{center}
\includegraphics[scale=0.3,angle=0] {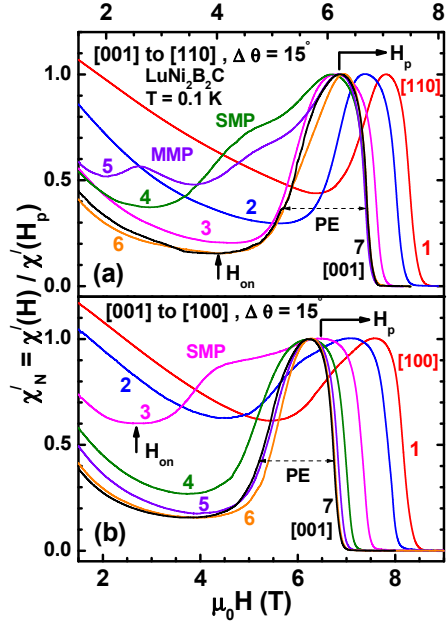}
\caption{ (Color online) Plots of normalised in-phase ac susceptibility ($\chi^{\prime}_N$) vs. applied field in $LuNi_2B_2C$ at T = 0.1~K, for change in orientation of the applied field from (a) [110] to [001] and (b) [100] to [001], in increments of $15^{\circ}$. }
\label{fig. 3}
\end{center}
\end{figure}

To elaborate more on anisotropy in effective disorder, we show in Fig.~3 plots of $\chi^{\prime}_N$ vs. applied field, for its movement out of the ab-basal plane in the LNBC crystal. Panels (a)/(b) show seven plots each for a change in orientation of the applied field from [110]/[100] (red curve) towards [001] direction (black curve), in steps of 15 degrees. The color coding and numerical labeling (1 to 7) identify the corresponding angles (measured w.r.t. [110]/[100]). Presence of a PE anomaly can be noted for all the orientations. As the field stands changed by $60^{\circ}$ from [110] towards [001] (cf. violet curve~-~5 in the panel (a)), one can witness multiple anomalies, termed as multiple magnetisation peaks (MMP), preceding the arrival of PE. However, such anomalies are not evident for the corresponding rotation from [100] in the panel (b) (cf. violet curve~-~5 in panel (b)). On changing the field-angle to $45^{\circ}$ in Fig.~3 (a)(cf. green curve~-~4), the PE anomaly remains considerably broadened and it can be imagined to be a juxtaposition of a SMP like anomaly preceding the PE. However, only a broad PE anomaly is observed for the corresponding rotation in Fig.~3~(b) (cf. green curves~-~4). As the orientation of the applied field stands changed by $30^{\circ}$ from [110]/[100] (cf. pink curves~-~3), only the PE anomaly is evident in the panel (a), but a SMP like anomaly now surfaces up in the panel (b). For the applied field changed by $15^{\circ}$ from [110]/[100] (cf. blue curves~-~2), the SMP anomaly can be imagined to move closer and juxtapose with PE in the panel (b), whereas only a PE is witnessed in the panel (a). For H $||$ [110]/[100] (cf. red curves~-~1), only a PE anomaly is observed, which as anticipated, is sharper in panel (a) vis-$\grave{a}$-vis panel (b).\\
SANS measurements on YNBC had revealed \cite{mckpaul}, that when the field was tilted $30^{\circ}$ away from the c-axis along the ac plane, a distorted hexagonal VL ($\beta \sim 75^{\circ}$) remained stabilised upto the highest measured field of 0.24~Tesla. In Fig.~3 (a), MMP anomalies are observed for such a $30^{\circ}$ change from [001] direction (cf. violet curve~-~5 in the panel (a)). Correspondence between SANS data in YNBC and our observation in Fig.~3 in LNBC could imply that when the applied field lies in a crystallographic direction that the symmetry of the VL is not close to (quasi)hexagonal with $\beta \sim 60^{\circ}$ or (quasi)square with $\beta \sim 90^{\circ}$, the spatial correlations in the VL could stand compromised as if there exists larger effective disorder. For field-angles such as $45^{\circ}$ w.r.t. [110] (curve~-~4 in Fig.~3 (a)) or $30^{\circ}$ w.r.t. [110] (curve~-~3 in Fig.~3 (b)), there are no SANS data in the literature, however, we anticipate that the VL symmetry, that results in SMP anomaly preceding the PE anomaly, is not high in such directions.
\begin{figure}[!htb]
\begin{center}
\includegraphics[scale=0.3,angle=270] {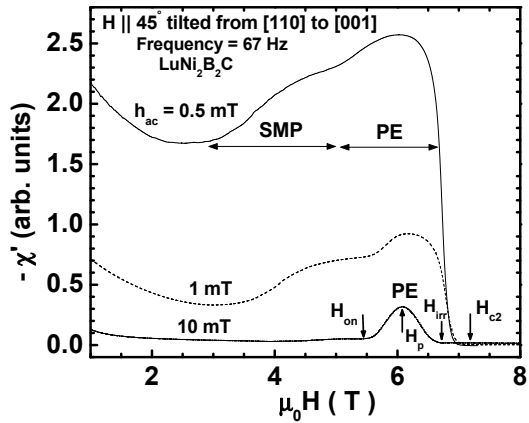}
\caption{Plots of (negative) in-phase ac susceptibility (-$\chi^{\prime}$) vs. applied field in $LuNi_2B_2C$ at T = 0.1~K, for H $||$ $45^{\circ}$ tilted from [110] towards [001], using an ac amplitude of 0.5~mT, 1~mT and 10~mT. The onset field ($H_{on}$), peak field ($H_p$), irreversibility field ($H_{irr}$) and upper critical field ($H_{c2}$) have been marked for $h_{ac}$~=~10~mT.}
\label{fig. 4}
\end{center}
\end{figure}

To demonstrate that SMP anomaly is, (purely) disorder induced, the disordered VL was shaken with a higher ac field to drive out the precursor SMP anomaly. Fig.~4 depicts the ac susceptibility data (-$\chi^{\prime}$ vs. field for different amplitudes, $h_{ac}$, of the ac field) at T~=~0.1~K for H corresponding to curve~-~4 in Fig.~3 (a). Note that for $h_{ac}$~=~0.5~mT (dashed line), and $h_{ac}$~=~1~mT (solid line), both SMP anomaly and PE anomaly are present. However, when the applied ac field is raised to 10~mT, only the PE anomaly can be witnessed. The ac field not only invokes the shielding currents in a superconductor, but also acts as a driving force, by attempting to make the metastable vortices reach their equilibrium configuration, while shaking them. The suppression of the SMP anomaly by shaking the VL establishes that the SMP anomaly is nucleated by an enhanced effective disorder.
\begin{figure}[!htb]
\begin{center}
\includegraphics[scale=0.3,angle=0] {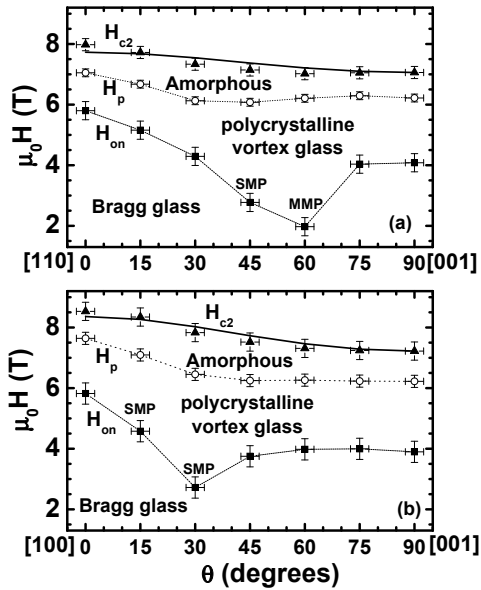}
\caption{The two panels show the variation of $H_{c2}$, $H_p$ and $H_{on}$, when the applied field is changed from (a) [110] to [001] and (b) [100]  to [001]. Dash and Dots are the guides to the eye, while the solid line is the fitting of $H_{c2}$($\theta$) to the Ginzburg-Landau theory. Various phases, e.g., Bragg glass, polycrystalline vortex glass and amorphous vortex state have been named, as per nomenclature in the literature \cite{g-l,satyajit}.}
\label{fig. 5}
\end{center}
\end{figure}

Collating  field-angles at which some typical (like, $H_{c2}$) as well as anomalous features (like, $H_{on}$ and $H_p$) could be noted in the ac susceptibility, we show in Fig.~5 the vortex phase diagrams for LNBC. The panels (a)/(b) in Fig.~5 correspond to the field tilted from [110]/[100] towards [001]. Dashes and dots are only guides to the eye, while the solid line is the fit to $H_{c2}$($\theta$) using the Ginzburg-Landau relationship \cite{tinkham}, $H_{c2} = H_{c2}(|| c,T)(sin^2(\theta) +\epsilon^2cos^2(\theta))^{-1/2}$, where $\epsilon = \frac{H_{c2}(||c)}{H_{c2}(||ab)}$, and $\theta$ is the angle between the applied field H and the ab-basal plane. From the fitting, we can obtain, $\epsilon$ $\approx$ 0.86 and 0.91, for field tilted from [110] and [100], respectively to [001], consistent with the earlier estimates \cite{me}. The average out of plane anisotropy, 0.5($H_{c2}^{[100]} + H_{c2}^{[110]})/H_{c2}^{[001]}$, is found to be 1.15, which also matches well with its value reported by Metlushko {\it et al.} \cite{metlushko}.\\
In Figs. 5~(a) and 5~(b), the field region is shown to be subdivided into three parts, (i) H $< H_{on}$ , (ii) $H_{on} <$ H $< H_p$ and H $> H_p$, and designated as well ordered Bragg glass (BG) state, polycrystalline vortex glass state and amorphous state, on the basis of notions articulated in the literature \cite{g-l,ling,satyajit}. The anisotropy in the phase diagram in Fig.~5 is striking, and may appear perplexing at the first glance. We rationalize a subtle difference between those two phase diagrams by recalling that the spatial order prior to $ H_{on}$ for H $||$ [110] is superior to that for H $||$ [100]. Thus, on progressively tilting the field from [110] towards [001], the field region of BG state gradually shrinks and reaches a minimum value at $\theta$ = $60^{\circ}$ (cf. Fig.~5(a)). On the other hand, starting from [100], the minimum is reached earlier at $\theta$ = $30^{\circ}$ (cf.~Fig.~5(b)) and, thereafter, the field interval of BG enhances as $H_{on}$ values climb up towards its plateau value for H $||$ [001].  Between $H_{on}$ and $H_p$, the vortex state is expected to comprise coexistence \cite{marchevsky} of ordered and disordered phases and above $H_p$, the disordered phase is the underlying equilibrium state, as had been shown for experiments in $V_3Si$ \cite{kupfer} and 2H-$NbSe_2$ \cite{marchevsky}. In the field ramping mode of experimentation, it has been shown \cite{paltiel1,paltiel2} that disordered bundles of vortices injecting through edges and corners continuously attempt to anneal to the underlying equilibrium vortex state, thereby, implying that residual disorder most likely coexists with the ordered state even prior to $H_{on}$. In platelet shaped sample, like the LNBC crystal, the extent of disorder injecting into the sample through edges and surfaces could depend on the orientation of the crystal, as has been witnessed while studying SMP phenomenon in crystals of high $T_c$ Bismuth cuprate, which had been cut in the form of a platelet and a prism \cite{kalisky}. While such a possibility can be a source of anisotropy in disorder/pinning in our case as well, we are more inclined to believe in the correlation between the anisotropy in disorder and the anisotropy in vortex lattice symmetry transition. The anisotropy in pinning observed between [110] and [100] in LNBC is analogous to that observed in a spherical crystal of (cubic) $V_3Si$ \cite{kupfer}. The shape dependent argument would not suffice to rationalize observations in spherical sample of $V_3Si$. \\

\section*{Conclusion}
The observation of large angular variation in the vortex phase diagram of a superconducting system, which {\it prime facie} imbibes marginal anisotropy in its intrinsic superconducting parameters, is a curious finding. Usually vortices are expected to assume triangular symmetry (apex angle $\beta$ = $60^{\circ}$) in the mixed state of a type-II superconductor, however, in quaternary borocarbide superconductors, $YNi_2B_2C$ and $LuNi_2B_2C$, the local symmetry of vortex lattice has been known to change from rhombohedral ($\beta < 60^{\circ}$) towards square ($\beta = 90^{\circ}$) at low fields. We note that for vortex arrays generated by applying field in different crystallographic directions of tetragonal structured (Y/L)NBC, the field interval over which good spatial order in the vortex lattice prevails correlates with the details of the underlying VL symmetry transition. For field oriented along crystal directions of higher symmetry, like, the c-axis or the ab-basal plane, where VL symmetry becomes perfect square or quasi-square, the well ordered Bragg glass state transforms to the disordered phase of vortex matter at high fields. Also, the c-axis, along which the VL assumes square symmetry, the correlation volume for the spatially ordered state is the largest and the shrinkage in the correlation volume across order-disorder transition {\it a la} peak effect is maximum. When the applied field is oriented away from c-axis towards the ab-basal plane, at intermediate angles ($\theta \sim 30^{\circ}$ to $60^{\circ}$), the field interval of Bragg glass state shrinks presumably due to enhancement in effective pinning. The fact that application of a large driving force can overcome the enhanced effective pinning and restore the field interval of Bragg glass phase supports the above assertion. Detailed field-angle dependent studies in suitably shaped (L/Y)NBC crystals are desired to fortify the correlation between anisotropy in VL symmetry transition and the angular variation in their vortex phase diagram. Experimental results presented will hopefully spur theoretical studies to quantitatively account for them.

\section*{Acknowledgments}
We thank Prof. M. R. Eskildsen and Prof. P. C. Canfield for few single crystals of $LuNi_2B_2C$, and Prof. B. Rosenstein for useful discussions. DJ-N acknowledges Sarojini Damodaran International Fellowship Programme for a visit to Tohoku University and Ministry of Education, Government of Japan, for a visiting scientist position.


\begin{thebibliography}{99}

\bibitem{blatter} BLATTER G., FEIGEL'MAN  M. V., GESHKENBEIN V. B.,  LARKIN A. I. and VINOKUR V. M., {\it Rev. Mod. Phys.}, {\bf 66} (1994) 1125.

\bibitem{g-l} GIAMARCHI T. and LE DOUSSAL P., {\it Phys. Rev. Lett.}, {\bf 72} (1994) 1530.

\bibitem{shobho} HIGGINS M. J. and BHATTACHARYA S., {\it Physica C}, {\bf 257} (1996) 232.

\bibitem{ling}  LING X. S., PARK S. R., McCLAIN B. A., CHOI S. M., DENDER D. C. and LYNN J. W., {\it Phys. Rev. Lett.}, {\bf 86} (2000) 712.

\bibitem{khaykovich} KHAYKOVICH B., ZELDOV E., MAJER D., LI T. W., KES P.H. and KONCZYKOWSKI M., {\it Phys. Rev. Lett.}, {\bf 76} (1996) 2555.

\bibitem{nishizaki} NISHIZAKI T. and KOBAYASHI N., {\it Supercond. Sci. Technol.}, {\bf 13} (2000) 1.

\bibitem{canfield} CANFIELD P. C., GAMMEL P. L. and BISHOP D. J., {\it Phys. Today}, {\bf 51} (1998) 40.

\bibitem{eskildsen1} ESKILDSEN M. R., GAMMEL P. L., BARBER B. P., RAMIREZ A. P., BISHOP D. J., ANDERSEN N. H., MORTENSEN K., BOLLE C. A., LIEBER C. M. and CANFIELD P. C., {\it Phys. Rev. Lett.}, {\bf 79} (1997) 487.

\bibitem{mckpaul} MCK. PAUL D., TOMY C. V., AEGERTER C. M., CUBITT R., LLOYD S. H., FORGAN E. M., LEE S. L. and YETHIRAJ M., {\it Phys. Rev. Lett.}, {\bf 80} (1998) 1517.

\bibitem{dewhurst} DEWHURST C. D., LEVETT S. J. and MCK. PAUL D., {\it Phys. Rev. B}, {\bf 72} (2005) 014542.

\bibitem{vinnikov1} VINNIKOV L. YA., BARKOV T. L., CANFIELD P. C., BUD'KO S. L., OSTENSON J. E., LAABS F. D. and KOGAN V. G., {\it Phys. Rev. B}, {\bf 64} (2001) 220508.

\bibitem{vinnikov2} VINNIKOV L. YA., BARKOV T. L., CANFIELD P. C., BUD'KO S. L. and KOGAN V. G., {\it Phys. Rev. B}, {\bf 64}(2001) 024504.

\bibitem{sakata} SAKATA H., OOSAWA M., MATSUBA K., NISHIDA N., TAKEYA H. and HIRATA K., {\it Phys. Rev. Lett.}, {\bf 84} (2000) 1583.

\bibitem{eskildsen2} ESKILDSEN M. R., ABRAHAMSEN A. B., LOPEZ D., GAMMEL P. L., BISHOP D. J., ANDERSEN N. H., MORTENSEN K. and CANFIELD P. C., {\it Phys. Rev. Lett.}, {\bf 86}(2001) 320-323.

\bibitem{takeya} TAKEYA H., HIRANO T.  and KADOWAKI K., {\it Physica C}, {\bf 256}, (1996) 220.

\bibitem{park} PARK T., CHIA E. E. M., SALAMON M. B., BAUER E. D., VEKHTER I., THOMPSON J. D., CHOI E. M., KIM H. J., LEE S.-I. and CANFIELD P. C., {\it Phys. Rev. Lett.}, {\bf 92}(2004) 237002.

\bibitem{kogan} KOGAN V. G., BULLOCK M., HARMON B., MIRANOVIC P., DOBROSAVLJEVIC–GRUJIC LJ., GAMMEL P. L. and BISHOP D. J., {\it Phys. Rev. B}, {\bf 55}(1997) R8693.

\bibitem{gammel} GAMMEL P. L., BISHOP D. J., ESKILDSEN M. R., MORTENSEN K., ANDERSEN N. H., FISHER I. R., CHEON K. O., CANFIELD P. C. and KOGAN V. G., {\it Phys. Rev. Lett.}, {\bf 82} (1999) 4082.

\bibitem{nakai} NAKAI N., MIRANOVIC P., ICHIOKA M., MACHIDA K., {\it Phys. Rev. Lett.}, {\bf 89} (2002) 237001.

\bibitem{bean} BEAN C. P., {\it Rev. Mod. Phys.}, {\bf 36} (1964) 31.

\bibitem{fietz} FIETZ W. A. and WEBB W. W., {\it Phys. Rev.}, {\bf 178} (1969) 657.

\bibitem{lo} LARKIN A. I. and OVCHINNIKOV YU. N., {\it J. Low Temp. Phys.}, {\bf 34} (1979) 409.

\bibitem{hc1} SCHMIEDESHOFF G. M., DETWILER J. A., BEYERMANN W. P., LACERDA A. H., CANFIELD P. C. and SMITH J. H., {\it Phys. Rev. B}, {\bf 63} (2001) 134519.

\bibitem{me} JAISWAL-NAGAR D., THAKUR A. D., ESKILDSEN M. R., CANFIELD P. C., YUSUF S. M., RAMAKRISHNAN S. and GROVER A. K., {\it Physica B}, {\bf 359-361}, (2005) 476.

\bibitem{metlushko} METLUSHKO V., WELP U., KOSHELEV A., ARANSON I., CRABTREE G. W. and CANFIELD P. C., {\it Phys. Rev. Lett}, {\bf 79} (1997) 1738.

\bibitem{kupfer} K$\ddot{U}$PFER H., LINKER G., RAVIKUMAR G., WOLF TH., WILL A., ZHUKOV A. A., MEIER-HIRMER R., OBST B. and WUHL H., {\it Phys. Rev. B}, {\bf 67} (2003) 064507.

\bibitem{toshi} ISSHIKI T., KIMURA N., AOKI H., TERASHIMA T., UJI S., YAMAUCHI K., HARIMA H., JAISWAL-NAGAR D., RAMAKRISHNAN S. and GROVER A. K., {\it Phys. Rev. B} {\bf 78} (2008) 134528.

\bibitem{kupfer2} K$\ddot{U}$PFER H., WOLF TH., ZHUKOV A. A. and MEIER-HIRMER R., {\it Phys. Rev. B}, {\bf 60} (1999) 7631.

\bibitem{budnick} LING X. S. and BUDNICK J. I., {\it Magnetic Susceptibilty of Superconductors and other Spin Systems}, Vol. {\bf 377} (Plenum Press) 1991.

\bibitem{satyajit} BANERJEE S. S., GROVER A. K., HIGGINS M. J., MENON G. I., MISHRA P. K., PAL D. RAMAKRISHNAN S., CHANDRASEKHAR RAO T. V., RAVIKUMAR G., SAHNI V. C., SARKAR S. and TOMY C. V., {\it Physica C}, {\bf 355}, (2001) 39.

\bibitem{tinkham} TINKHAM M., {\it Introduction to Superconductivity}, Mc Graw-Hill Inc. (New York) 1996, Second edition.

\bibitem{marchevsky} MARCHEVSKY M., HIGGINS M. J. and BHATTACHARYA S., {\it Nature}, {\bf 409} (2001) 591.

\bibitem{paltiel1} PALTIEL Y., ZELDOV E., MYASOEDOV Y. N., SHTRIKMAN H., BHATTACHARYA S., HIGGINS M. J., XIAO Z. L., ANDREI E. Y., GAMMEL P. L. and BISHOP D. J., {\it Nature}, {\bf 403} (2000) 398.

\bibitem{paltiel2} PALTIEL Y., ZELDOV E., MYASOEDOV Y., RAPPARORT M. L., JUNG G., BHATTACHARYA S., HIGGINS M. J., XIAO Z. L., ANDREI E. Y., GAMMEL P. L. and BISHOP D. J., {\it Phys. Rev. Lett}, {\bf 85} (2000) 3712.

\bibitem{kalisky} KALISKY B., MYASOEDOV Y., SHAULOV A., TAMEGAI T., ZELDOV E. and YESHURUN Y., {\it Phys. Rev. Lett}, {\bf 98} (2007) 107001.

\end{thebibliography}
\end{document}